\documentstyle[12pt]{article}
\begin{document}
\baselineskip 22pt

\vspace*{.3in}

\begin{center}
{\Large UNCONDITIONAL SECURITY \\
IN QUANTUM BIT COMMITMENT} \\
\vspace*{.4in}

{\Large Horace P. Yuen} \\ 
{\large Department of Electrical and Computer
Engineering \\ Department of Physics and Astronomy\\
Northwestern University \\ Evanston IL  60208-3118 \\ email: yuen@ece.northwestern.edu}
\end{center}
\vspace*{.4in}

\begin{abstract}
The commitment of bits between two mutually distrustful parties is a powerful cryptographic primitive with which many cryptographic objectives can be achieved. It is widely believed that
unconditionally secure quantum bit commitment is impossible due to
quantum entanglement cheating, which is codified in a general
impossibility theorem. Gaps in the proof of this impossibility theorem are found.  An unconditionally secure bit
commitment protocol utilizing anonymous quantum states and the no-clone theorem is presented
below with a full security proof.

\vspace*{.2in}

\noindent PACS \#: 03.67Dd, 03.65Bz
\end{abstract}

\vspace*{.4in}

\baselineskip 11pt

NOTE:

\begin{enumerate}

\item This paper is self-contained.  The titles of this paper and {\tt
quant-ph/0006109} [referred to as (I) in this note] will be
interchanged when a revision of (I) is completed.  In (I), three QBC
protocols (QBC1,2,3) are described that may be unconditionally secure.

\item Protocol QBC1 can indeed be proved secure, as will be shown in
the revision of (I).

\item Protocol QBC2, as presented in (I), has a security gap that can
be filled in various ways.  One way is to use decoy states as
described in this paper.

\item Protocol QBC3 in (I) is insecure, similar to QBC01 also described in
(I).  The name QBC3 is taken over by the protocol of this paper, which
may be viewed as a simplification of QBC2.

\item In the forthcoming revision of (I) and in this paper, the places
where the impossibility proof fails in each of the protocols will be
precisely pinpointed.

\end{enumerate}

\newpage

\baselineskip 22pt

Quantum cryptography \cite{bennett}, the study of information security
systems involving quantum effects, has recently been associated almost
exclusively with the cryptographic objective of key distribution.
This is due primarily to the nearly universal acceptance of the general
impossibility of secure quantum bit commitment (QBC), taken to be a
consequence of the Einstein-Podolsky-Rosen (EPR) type entanglement
cheating which rules out QBC and other quantum protocols that have
been proposed for various other cryptographic objectives
\cite{brassard}.  In a bit commitment scheme, one party, Adam,
provides another party, Babe, with a piece of evidence that he has
chosen a bit b (0 or 1) which is committed to her.  Later, Adam would
``open'' the commitment: revealing the bit b to Babe and convincing
her that it is indeed the committed bit with the evidence in her
possession.  The usual concrete example is for Adam to write down the
bit on a piece of paper which is then locked in a safe to be given to
Babe, while keeping for himself the safe key that can be presented
later to open the commitment.  The evidence should be {\em binding},
i.e., Adam should not be able to change it, and hence the bit, after
it is given to Babe.  It should also be {\em concealing}, i.e., Babe
should not be able to tell from it what the bit b is.  Otherwise,
either Adam or Babe would be able to cheat successfully.

In standard cryptography, secure bit commitment is to be achieved
either through a trusted third party or by invoking an unproved
assumption on the complexity of certain computational problem.  By
utilizing quantum effects, various QBC schemes not involving a third
party have been proposed that were supposed to be unconditionally
secure, in the sense that neither Adam nor Babe can cheat with any
significant probability of success as a matter of physical laws.  In
1995-1997, a general proof on the impossibility of unconditionally
secure QBC and the insecurity of previously proposed protocols were
described \cite{mayers1}-\cite{lo}.  Henceforth, it has been accepted
that secure QBC and related objectives are impossible as a matter of
principle \cite{lo2}-\cite{brassard2}.

Since there is no known characterization of all possible QBC
protocols (or indeed all possible cryptographic protocols of any kind)
with corresponding performance characterization, logically there can
really be no general impossibility proof even if it were indeed
impossible to have a secure QBC protocol. In this paper, a QBC scheme
utilizing anonymous states and decoy states will be presented with an unconditional security proof.  The basic reason for its success is that the flow of classical information between the two parties is not properly accounted for in the impossibility proof. The results are developed within nonrelativistic quantum mechanics, unrelated to relativistic protocols \cite{kent}. The QBC framework is as follows.

When Adam picks b = 0 to commit to Babe, he sends her a
state $|\phi _i \rangle \in {\cal H}^B$ with probability $p_i$ within
fixed openly known sets $\{ |\phi _i \rangle \}$ and $\{ p_i \}$ for
$i \in \{ 1, \cdots , M \}$.  When he picks b = 1, he sends $|\phi '
_i \rangle \in {\cal H}^B$ from another fixed openly known set $\{
|\phi ' _i \rangle \}$ with probabilities $\{ p' _i \}$. The $\{ |\phi _i \rangle \}$ and $\{
\phi ' _i \rangle \}$ are so chosen that they are concealing as
evidence, i.e., Babe cannot reliably discriminate between them in
optimum binary quantum hypothesis testing \cite{helstrom}.  They would
also be binding if Adam is honest and sends them as they are, which he
could not change after Babe receives them.  In that case, when Adam
reveals the bit by telling exactly which state $|\phi _i \rangle$ or
$|\phi ' _i \rangle$ he sent, Babe can measure the corresponding
projector to verify the bit.  In general, Babe can always guess the
bit with a probability of success $P^B_c = \frac{1}{2}$, while Adam
should not be able to change a committed bit at all.  However, it is
meaningful and common to grant {\em unconditional security} when the
best $\bar{P}^B_c$ Babe can achieve is arbitrarily close to 1/2 and
Adam's best probability of successfully changing a committed bit
$\bar{P}^A_c$ is arbitrarily close to zero \cite{mayers} even when
both parties have perfect technology and unlimited resources including
computational power. 

The impossibility proof gives the following general EPR
cheat that Adam can launch. Instead of
sending $|\phi _i \rangle$ or $|\phi' _i \rangle$, Adam can generate
$|\Phi _0 \rangle$ or $|\Phi _1 \rangle$ depending on b = 0 or 1,
\begin{equation}
|\Phi _0 \rangle = \sum_i \sqrt{p_i} | e_i \rangle | \phi _i
 \rangle , \hspace*{.2in}
 |\Phi _1 \rangle = \sum_i \sqrt{p'_i} | e'_i \rangle | \phi ' _i \rangle
\end{equation}
with $\{ | e_i \rangle \}$, $\{ | e'_i \rangle \}$ complete
orthonormal in ${\cal H}^A$, and sends Babe ${\cal H}^B$ while keeping
${\cal H}^A$ himself.  He can switch between $|\Phi _0
\rangle$ and $|\Phi _1 \rangle$ by operation on ${\cal H}^A$ alone,
and thus alter the evidence to suit his choice of b before opening the
commitment.  In the case $\rho^B_0 \equiv tr_A |\Phi _0 \rangle
\langle \Phi _0 | = \rho ^B _1 \equiv tr_A |\Phi _1 \rangle \langle
\Phi _1|$, the switching operation is to be obtained by using the
so-called ``Schmidt decomposition,'' the expansion of
$|\Phi _0 \rangle$ and $|\Phi _1 \rangle$ in terms of the eigenstates
$|\hat{\phi}_k \rangle$ of $\rho^B_0 = \rho^B_1$ and the eigenstates
$|\hat{e}_k \rangle$ and $|\hat{e}'_k \rangle$ of $\rho^A_0$ and
$\rho^A_1$,
\begin{equation}
|\Phi _0 \rangle = \sum _k \lambda^{1/2}_k |\hat{e}_k \rangle
 |\hat{\phi}_k \rangle, \hspace*{.2in}
|\Phi _1 \rangle = \sum _k \lambda^{1/2}_k |\hat{e}'_k \rangle
 |\hat{\phi}_k \rangle
\end{equation}
By applying a unitary $U^A$ that brings $\{ |\hat{e}_k \rangle \}$ to
$\{ |\hat{e}'_k \rangle \}$, Adam can select between $|\Phi_0 \rangle$
or $|\Phi_1 \rangle$ any time before he opens the commitment but after
he supposedly commits.  When $\rho_0^B$ and $\rho_1^B$ are not equal
but close, it was shown that one may transform $| \Phi _0 \rangle$ by an $U^A$ to a
$| \tilde{\Phi} _0 \rangle$ with $|\langle \Phi _1 | \tilde{\Phi} _0 \rangle |$ as
close to 1 as $\rho ^B_0$ is close to $\rho ^B_1$ according to the
fidelity F chosen, and thus the state $| \tilde{\Phi} _0 \rangle$ would serve
as the effective EPR cheat.  This $U^A$ is determined from knowledge of $p_i,  p'_i, \phi_i$, and $\phi '_i$ \cite{jozsa}.

Using the anonymous quantum state technique \cite{yuen}, the bit
commitment may proceed as follows.  Babe transmits to Adam a state $| \psi \rangle$ only known to herself.  Adam sends her back a committed bit b via modulating $|\psi\rangle$ by openly known unitary operators $U_{\rm b}$.  Thus, the states $|\phi_i\rangle$ and $|\phi'_i\rangle$ in (1) are not known to Adam, and the corresponding cheating $U^A$ cannot be found in the case $\rho^B_0$ is close, but not identical to, $\rho^B_1$ \cite{note13}.  How is the impossibility proof supposed to work in this case?  It appears from \cite{mayers}, and especially \cite{lo1}, that such classical randomness introduced in the protocol is to be turned into quantum determinateness via quantum entanglement purification of a mixed quantum state.  Such a prescription fails to preserve the protocol for several reasons \cite{note13}, one of which is exploited in the protocol QBC3 to be described in this paper.  Generally, the impossibility proof appears to suffer from the glaring and severe scope problem --- why are all possible QBC protocols covered by its transparently very specific formulation?  For example, why are $\rho^B_0$ and $\rho^B_1$ necessarily the marginal states obtained by tracing over the states generated by Adam as in (1)?  Protocol QBC3 shows clearly that this is not the case.  A precise specification of QBC3 with a full security proof will be given after the following description of how the protocol works.  

Let the state $| \psi \rangle$ sent by Babe be an arbitrary state of a
qubit, a two-dimensional quantum state space.  Thus it is described by
a three-dimensional unit vector on the Bloch-Poincar\'{e} sphere
\cite{nielchu}.  Let $U_0 = I$ and $U_1 = R(\theta,C)$, a rotation by
an angle $\theta$ on some great circle $C$ on the sphere.  To fix
ideas, we may let $\theta = \pi$, so that $U_0|\psi\rangle$ and
$U_1|\psi\rangle$ are orthogonal when $| \psi \rangle$ is on $C$, but
this choice is not mandatory.  It is clear that Adam cannot cheat in
this case, since there is only one possible state for each bit value,
and so no possibility for entanglement --- assuming, as in the
impossibility proof, that he is going to maintain a perfect opening
for ${\rm b}=0$, so that entanglement cheating is to be used for
opening ${\rm b}=1$.  The general case will be dealt with later.  On
the other hand, Babe can cheat perfectly by measuring the basis $\{ |
\psi \rangle, R(\pi,C)|\psi\rangle \}$ with $| \psi \rangle$ on $C$.
If the two states $U_0 |\psi\rangle$ and $U_1|\psi\rangle$ are chosen
to be close to make $\bar{P}^B_c - 1/2$ small, $\bar{P}^A_c$ is close
to one by simply declaring ${\rm b}=1$ on the committed
$U_0|\psi\rangle$.  Thus, we maintain the above $U_{\rm b}$ and defeat
Babe's cheating by the use of {\em decoy states}. Instead of just one qubit $U_{\rm b} |\psi\rangle$, Adam can send back to Babe a sequence of $n$ qubits
\begin{equation}
| \psi_ 1 \rangle _1 \, \ldots \, |\psi _i \rangle _i \, \ldots \, |\psi _n  \rangle _n,
\label{qubitseq}
\end{equation}
each named by its position in the sequence --- e.g., state $i$ describes the qubit occupying the nonoverlapping $i$th time interval.  In (\ref{qubitseq}), one of the $|\psi _i \rangle$ is randomly chosen to be $U_{\rm b} | \psi \rangle$, the others are independently and randomly chosen to be arbitrary qubit states.  (The orientation of each Bloch sphere for each qubit is, as usual, assumed known to both parties).  Adam opens by telling Babe which $| \psi _ i \rangle$ is $U_{\rm b} |\psi\rangle$ and what ${\rm b}$ is.

Since Adam still cannot cheat via entanglement --- he needs to
identify the exact qubit (indeed local state invariance \cite{note13}
would be violated if he could cheat via entanglement) --- it is clear
that he can only cheat with a fixed (not arbitrarily small)
probability; for $|\psi\rangle$ on $C$ it is $\bar{P}^A_c = 1/2$,
corresponding to the probability that a randomly chosen state on $C$
could be accepted as a fixed given state on $C$.  This cheating
probability $1/2$ is obtained by announcing another $|\psi_i\rangle
\neq U_0|\psi\rangle$ to be $U_1|\psi\rangle$. With perfect
opening on ${\rm b}=0$, the no-clone
theorem \cite{wz,yuen_noclone} prevents Adam from doing any better; it
makes the use of decoy states secure and demonstrates the quantum
nature of the protocol.  Note that the use of anonymous states from
Babe is essential for preventing Adam's cheating.  If the condition of
perfect opening on ${\rm b}=0$ is relaxed, Adam can employ an optimal
one-to-two cloner on $|\psi\rangle$, apply $U_0$ and $U_1$ to them,
and open accordingly.  The optimal $\bar{P}^A_c$ in this case is again
some fixed number $\bar{p}$, not arbitrarily close to one.  On the other hand, Babe's cheating probability can be made arbitrarily small by having $n$ large.  This is intuitively obvious because, in order to
cheat with $\bar{P}^B_c > 1/2$, Babe needs to either guess correctly
the $i$th position that carries $U_{\rm b}|\psi\rangle$ or to take a
majority vote from measurement results among the $n$ qubits with only
the advantage of one qubit in her favor.  It turns out that this
second strategy is optimum for her, but evidently it still performs
poorly.  Her entanglement would not help because she is trying to
decide, rather than to change, what she has sent. Indeed, the
following proof shows that even if she adjoins her entangled qubit
positions correctly, the entanglement does not improve $\bar{P}^B_c$.
Thus we now have a situation directly contradicting the strong claim of the impossibility proof, which goes beyond the mere impossibility of unconditional security by asserting that whenever $\bar{P}^B_c - 1/2$ is arbitrarily small, $\bar{P}^A_c$ is arbitrarily close to one.

At this point, it is appropriate to examine why the impossibility
proof fails to work in this basically very simple protocol.  One way
to look at it, as indicated above, is that $\rho^B_0$ and $\rho^B_1$
are not the marginal states obtained by tracing over $|e_i\rangle$
and $|e'_i\rangle$ in $|\Phi_0\rangle$ and $|\Phi_1\rangle$, due to the introduction of additional qubit position
randomness.  As an alternative way to discern this failure of the
impossibility proof, one may understand the impossibility proof as
first granting the condition $\rho^B_0$ is close to $\rho^B_1$ and
then purporting to show that $\bar{P}^A_c$ is close to 1 as a
consequence, with the change of classical randomness into quantum
determinateness.  In QBC3, this would entail that Babe would generate
an entangled state $\sum_i \sqrt{\lambda_i} |\psi_i\rangle_B
|f_i\rangle_C$ with openly known $\lambda_i,|\psi_i\rangle$ and send
${\cal H}^B$ to Adam, while keeping ${\cal H}^C$ to herself.  (Whether
$i$ is from a continuum or a finite set is not important here).  At
the end of the commitment phase she would measure the orthogonal basis
$\{ |f_i\rangle \}$.  An error of the impossibility proof can now be seen.  For her verification in QBC3, Babe only accepts one state
$|\psi_i\rangle$, the state finally sent, and thus is {\em using} her
measurement result, while the measurement result for such classical
randomness is implicitly assumed {\em not} to be used in the
impossibility proof, which in fact does not carry any description of
the utilization of such measurement results.  It is precisely such
total lack of any role for classical information flow and utilization
between the two parties that makes the impossibility proof severely
limited in scope, and incorrect as a general proof.  This problem also manifests itself in opening other gaps in the impossibility proof, see \cite{note13} for further discussion.

While the above development shows how and why the impossibility proof
breaks down, it does not yet show that unconditionally secure quantum
bit commitment is possible because $\bar{P}^A_c$ is not arbitrarily
small.  However, it can be made so in a sequence rather than a single qubit.  Let Babe send Adam a sequence of $m$ qubits
\begin{equation}
|\psi^1\rangle_1 \, \ldots \, |\psi^j\rangle_j \, \ldots \,|\psi^m\rangle_m,
\label{qubitseq2}
\end{equation}
each randomly and independently chosen from $C$ and named by its
temporal position $j$.  Depending on whether ${\rm b}=0$ or ${\rm
b}=1$, Adam applies $U_0$ or $U_1$ to each of these $m$ qubits,
randomly place each $|\psi^j\rangle_j$ in a sequence of $n$ qubit
states (\ref{qubitseq}), each newly named by its temporal position $i$
in the $n$-sequence (\ref{qubitseq}) when sent back to Babe for
commitment, with the other $n-m$ being independently and randomly
chosen arbitrary qubit states.  Thus, each of the $(n)_m = n!/(n-m)!$
ordered $m$-qubit positions in the $n$ sequence has the same
probability $\frac{1}{(n)_m}$ of being chosen to accommodate the
modulated $m$ qubit states (\ref{qubitseq2}).  Adam opens by telling
Babe the bit value and which $|\psi_i\rangle_i$ is $U_{\rm
b}|\psi^j\rangle_j$, for all $j$.  It is clear that Adam's
$\bar{P}^A_c$ is changed from the single-qubit value $\bar{p}$ to
$\bar{p}^m$, while $\bar{P}^B_c-1/2$ can still be made arbitrarily
small when $n$ is sufficiently large.  Again, Adam has no entanglement
attack, and Babe's entanglement serves no useful purpose.  A precise proof follows.\\

\noindent{PROTOCOL QBC3}.

\begin{enumerate}

\item[(i)] Babe sends Adam a sequence of states (\ref{qubitseq2}), each independently and randomly chosen from a fixed great circle $C$ on the Bloch sphere of each of the $m$ qubits; 

\item[(ii)] Adam modulates each of these states either by $U_0$ given by the identity transformation $I$ or by $U_1$ given by the rotation by $\pi$ radians on $C$, according to ${\rm b}=0$ or ${\rm b}=1$.  He then independently and randomly places them among $n$ qubits positions of (\ref{qubitseq}) and picks arbitrary states on $C$ for the other $n-m$ qubits, sending the $n$ qubits to Babe as commitment;

\item[(iii)] Adam opens by revealing which position each $U_{\rm b}|\psi^j\rangle_j$ from (\ref{qubitseq2}) takes in the $n$-sequence (\ref{qubitseq}), and the bit value.  Babe verifies by measuring the corresponding projections.

\end{enumerate}

\noindent{To see that $\bar{P}^A_c$ can be made arbitrarily small, we have already showed that Adam has no entanglement cheating since he has to identify each individual qubit and there is only one states associated with each bit value for each qubit.  Optimal one-to-two cloner for each qubit from (\ref{qubitseq2}) can be employed, with an appropriate criterion given by the average over individual qubit inner products that yields $\bar{p}$ described above. While there are many results in the literature on approximate cloning \cite{kwclone}, the most general one appropriate to our criterion $\bar{p}$ does not seem to have been worked out.  Nevertheless, $\bar{p}$ is some fixed number, so that $\bar{P}^A_c = \bar{p}^m$ can be made arbitrarily small in an $m$-sequence.}

To show that $\bar{P}^B_c - 1/2$ can also be made arbitrarily small, first consider the $m=1$ case with no entanglement to illustrate the main idea.  In this case, Babe's density operators are, for ${\rm b}=0,1$,
\begin{equation}
\rho^B_{\rm b} = \frac{1}{n} \sum_i \frac{I}{2} \otimes \ldots \otimes \stackrel{i}{\sigma}_{\rm b} \otimes \ldots \otimes \frac{I}{2},
\label{babe_do}
\end{equation}
where $\sigma_{\rm b}$, occurring in each of the $n$
positions with equal probability, is the qubit state $U_{\rm b}\sigma
U^\dag_{\rm b}$ modulated by Adam
with $\sigma = |\psi\rangle \langle\psi|$ if Babe sends a pure state.  Since \cite{helstrom,fuchs}
\begin{equation}
\bar{P}^B_c - \frac{1}{2} = \frac{1}{4} \| \rho^B_0 - \rho^B_1 \|_1,
\label{trnormest}
\end{equation}
where $\| \cdot \|_1$ is the trace norm \cite{nielchu,fuchs}, and $\rho^B_0 - \rho^B_1$ from (\ref{babe_do}) is diagonal in the product basis that diagonalizes $\sigma_0 - \sigma_1$ on each qubit, one finds straightforwardly from (\ref{trnormest}) that, assuming $n = 2l+1$,
\begin{equation}
\bar{P}^B_c - \frac{1}{2} = \frac{\lambda_+}{2^n}{2l \choose l} < \frac{1}{2\sqrt{\pi l}},
\label{barpcest}
\end{equation}
where $\lambda_+ \le 1$ is the positive eigenvalue of $\sigma_0 - \sigma_1$.  The optimal probability in (\ref{barpcest}) can be obtained by measuring in the above product basis and setting ${\rm b}=0$ or $1$ according to a majority vote on the positive and negative ouctomes (corresponding to the eigenvectors $|\lambda_+\rangle$ and $|\lambda_-\rangle$).  Since (\ref{barpcest}) also implies
\begin{equation}
\bar{P}^B_c - \frac{1}{2} > \frac{1}{4\sqrt{l}},
\label{barpbc_lb}
\end{equation}
the optimal strategy is better than guessing at the qubit sent, which yields $\bar{P}^B_c = \frac{1}{2}\left(1+\frac{1}{n}\right)$.

Intuitively, because Babe does not know the positions of the qubits she sent, in the general $m$ case with entanglement her optimal strategy for $\bar{P}^B_c$ is to send all the $|\psi^j\rangle_j$ of (\ref{qubitseq2})  in the same uniform state $|\psi\rangle$ without entanglement, measure the diagonal basis of $U_0|\psi\rangle \langle \psi|U^\dag_0- U_1 |\psi\rangle \langle \psi|U^\dag_1$ for each qubit from (\ref{qubitseq}), and take a majority vote.  For $n = m + 2l' = 2l+1$, the resulting
\begin{equation}
\bar{P}^B_c - \frac{1}{2} = \frac{1}{2^{n-m+1}} {n-m \choose l'} + \frac{1}{2^{n-m}} \sum^l_{k=l'+1} {n-m \choose k}
\label{barpbcest2}
\end{equation}
is bounded above by $(m+1)/(2\sqrt{\pi l'})$ that goes to zero for large $l' = (n-m)/2$.  To handle her possible entanglement on (\ref{qubitseq2}), it is simpler to proceed as follows instead of using (\ref{trnormest}) directly.

To use her entanglement, Babe needs to pick $m$ out of $n$ qubit
positions, to which she would adjoin her entangled qubits for
measurement.  For each choice, the probability that none of these $m$
positions would overlap with any of the qubits she sent in the
$n$-sequence (\ref{qubitseq}) is given by the hypergeometric
distribution $\frac{{n-m \choose m}}{{n \choose m}}$ which is
arbitrarily close to 1 for $m/n$ sufficiently small.  In this case, her part of the entangled state drops out and the value $\| \rho^B_0 - \rho^B_1 \|_1$ is given arbitrarily closely, explicitly proved by applying the triangle inequality for the trace norm and noting that $\| \rho_0 - \rho_1 \|_1 \le 2$ for any two density operators, through the following $\rho^B_{\rm b}$ with no entanglement:
\begin{equation}
\rho^B_{\rm b} = \frac{1}{(n)_m} \sum_{j \in J} \frac{I}{2} \otimes \ldots \otimes \stackrel{j}{\sigma}_{m{\rm b}} \otimes \ldots \otimes \frac{I}{2},
\label{babe_do2}
\end{equation}
where $J$ is the set of $(n)_m$ $m$-positions out of $n$, and
$\sigma_{m{\rm b}}$ denotes the joint $m$-qubit state
for any of the $j$th $m$-position \cite{note19}.  Since any qubit
measurement statistics can be obtained by a product basis and joint
state across $m$ qubits, it follows that the optimal $\bar{P}^B_c$
from (\ref{babe_do2}) can be obtained from a classical joint
distribution on individual qubit variable measurements.  The resulting
optimal quantum performance cannot be better than that of optimizing a
classical joint distribution or the corresponding Kolmogorov distance
\cite{nielchu,fuchs}, because there is no compatible basis issue in
the classical case. A direct term-by-term bounding of the Kolmogorov
distance shows that the optimal classical solution indeed corresponds
to the independent uniform-state qubits one indicated above, which is
expected if only because Babe cannot effectively use any correlation
between any of the $m$ qubits without knowing their positions.  Note
that Babe is actually not advised to send uniform qubit states to Adam, who can then clone much better in an $m$-to-$2m$ approximate cloner.  

Protocol QBC3 achieves unconditional security with the use of
anonymous states to thwart Adam's cheating and decoy states to
thwart Babe's cheating.  In \cite{note13}, it will be shown that
anonymous states alone without decoy states could lead to
unconditional security in a more complicated protocol QBC1, in which
Babe's cheating is thwarted via bit hashing.

\vspace*{.2in}

\begin{center}
ACKNOWLEDGMENT
\end{center}

\vspace*{.2in}

I would like to thank B. Leslau and J. Mueller-Quade for useful discussions. This work was supported in part by the Defense Advanced Research Project Agency and in part by the Army Research Office.


\begin{thebibliography}{99}
\bibitem{note19}The irrelevance of Babe's entanglement can also be
observed from the fact that even if one grants that the adjoined
positions of the entangled qubits are correct for each possible $m$-qubit position
in $J$, one is still left with $\rho^B_{\rm b}$ of the form (\ref{babe_do2}).
\end{thebibliography}
\end{document}